\documentclass[longauth,traditabstract]{aa} % for the abstract without structuration 
\usepackage{graphicx}
\usepackage{natbib}
\usepackage{txfonts}
\begin{document}
\title{Noise properties of the CoRoT\thanks{ The CoRoT space
    mission, launched on December 27th 2006, has been developed and is
    operated by CNES, with the contribution of Austria, Belgium,
    Brazil, ESA, Germany, and Spain. CoRoT data become publicly
    available one year after release to the Co-Is of the mission from
    the CoRoT archive: {\tt http://idoc-corot.ias.u-psud.fr/}.} data}
\subtitle{A planet-finding perspective}

   \author{S.\ Aigrain\inst{1} \and % Exeter
     F.\ Pont\inst{1} \and % Exeter
     F.\ Fressin\inst{2} \and % CfA
     A.\ Alapini\inst{1} \and %
     R.\ Alonso\inst{3} \and % LAM
     M.\ Auvergne\inst{4} \and % LESIA
     M.\ Barbieri\inst{3} \and % LAM
     P.\ Barge\inst{3} \and % LAM
     P.\ Bord{\' e}\inst{5} \and % IAS
     F.\ Bouchy\inst{6} \and % OHP
     H.\ Deeg\inst{7} \and % IAC
     R.\ De la Reza\inst{8} \and % ON Brazil
     M.\ Deleuil\inst{3} \and % LAM 
     R.\ Dvorak\inst{9} \and % Vienna
     A.\ Erikson\inst{10} \and % DLR
     M.\ Fridlund\inst{11} \and % ESTEC
     P.\ Gondoin\inst{11} \and % ESTEC
     P.\ Guterman\inst{3} \and % LAM
     L.\ Jorda\inst{3} \and % LAM
     H.\ Lammer\inst{12} \and % AAS
     A.\ L{\' e}ger\inst{5} \and % IAS
     A.\ Llebaria\inst{3} \and % LAM
     P.\ Magain\inst{13} \and % ULG
     T.\ Mazeh\inst{14} \and % Tel Aviv
     C.\ Moutou\inst{3} \and % LAM
     M.\ Ollivier\inst{5} \and % IAS
     M.\ Paetzold\inst{15} \and % Koln
     D.\ Queloz\inst{16} \and % Geneva
     H.\ Rauer\inst{10,17} \and % DLR
     D.\ Rouan\inst{4} \and % LESIA
     J.\ Schneider\inst{18} \and % LUTH
     G.\ Wuchter\inst{19} \and % Tautenburg
     S.\ Zucker\inst{20} % Tel Aviv 2
}

          \institute{
            School of Physics, University of Exeter, Exeter, 
            EX4 4QL, UK \and
            Harvard-Smithsonian Center for Astrophysics, 60 Garden Street,
            Cambridge, MA 02138, US \and \ldots
            LAM, Universit{\' e} de Provence, 13388
            Marseille, France \and
            LESIA, Observatoire de Paris, 92195 Meudon,
            France \and
            IAS, Universit{\' e} Paris XI, 91405 Orsay, France \and
            Observatoire de Haute-Provence, 04870 St Michel l’Observatoire, 
            France \and 
            IAC, E-38205 La Laguna, Spain \and
            ON/MCT, 20921-030, Rio de Janeiro, Brazil \and
            IfA, University of Vienna, 1180 Vienna, Austria \and
            Institute of Planetary Research, DLR, 12489 Berlin,
            Germany \and
            RSSD, ESA/ESTEC, 2200 Noordwijk, The
            Netherlands \and
            IWF, Austrian Academy of Sciences,
            A-8042 Graz, Austria \and
            IAG, Universit{\' e} de Li{\` e}ge, Li{\` e}ge 1, Belgium
            \and
            Sch.\ Physics \& Astronomy, Tel Aviv Univ., Tel
            Aviv 69978, Israel \and
            RIU,
            Universit{\"a}t zu K{\" o}ln, 50931 K{\" o}ln, Germany \and
            Observatoire de Gen{\` e}ve, 1290 Sauverny, Switzerland
            \and
            ZAA, TU Berlin,
            D-10623 Berlin, Germany \and
            LUTH, Observatoire de Paris, 92195 Meudon, France \and
            Th{\" u}ringer Landessternwarte, 07778
            Tautenburg, Germany \and
            Dept.\ Geophysics \& Planetary Sciences, Tel Aviv
            Univ., Tel Aviv 69978, Israel
        }

   \date{Received \ldots; accepted \ldots}

   \abstract{In this short paper, we study the photometric precision
     of stellar light curves obtained by the CoRoT satellite in its
     planet finding channel, with a particular emphasis on the
     timescales characteristic of planetary transits. Together with
     other articles in the same issue of this journal, it forms an
     attempt to provide the building blocks for a statistical
     interpretation of the CoRoT planet and eclipsing binary catch to
     date.

     After pre-processing the light curves so as to minimise long-term
     variations and outliers, we measure the scatter of the light
     curves in the first three CoRoT runs lasting more than 1 month,
     using an iterative non-linear filter to isolate signal on the
     timescales of interest. The bevhaiour of the noise on 2\,h
     timescales is well-described a power-law with index 0.25 in
     $R$-magnitude, ranging from 0.1\,mmag at $R=11.5$ to 1\,mmag at
     $R=16$, which is close to the pre-launch specification, though
     still a factor 2--3 above the photon noise due to residual jitter
     noise and hot pixel events. There is evidence for a slight
     degradation of the performance over time. We find clear evidence
     for enhanced variability on hours timescales (at the level of
     0.5\,mmag) in stars identified as likely giants from their $R$
     magnitude and $B-V$ colour, which represent approximately 60 and
     20\% of the observed population in the direction of Aquila and
     Monoceros respectively. On the other hand, median correlated
     noise levels over 2\,h for dwarf stars are extremely low,
     reaching 0.05\,mmag at the bright end.}


   \maketitle
%
%________________________________________________________________

\section{Introduction}

\begin{figure*}
  \centering
  \includegraphics[width=\linewidth]{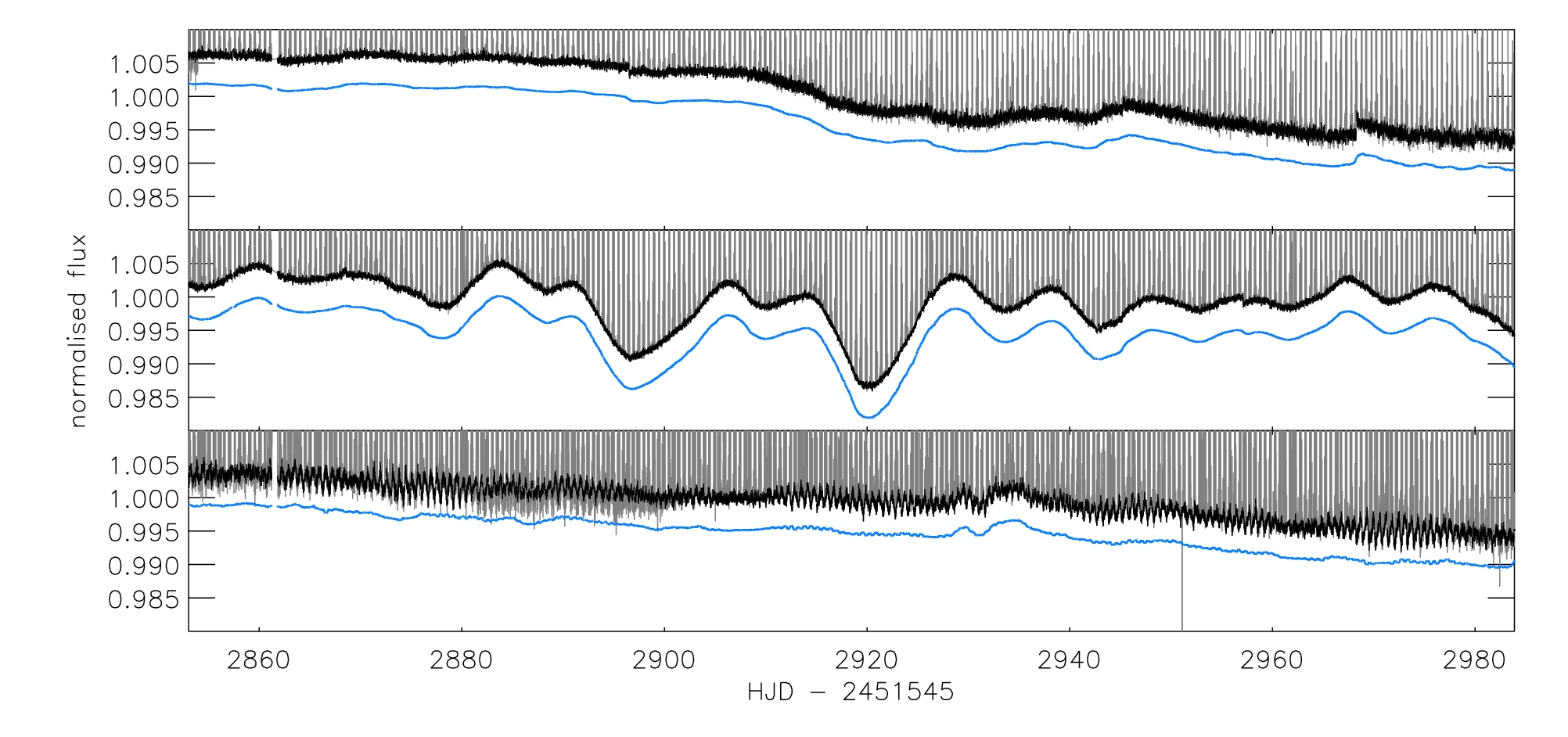}
  \caption{Example light curves from run LRa01 before and after
    clipping (grey and black respectively). The variations on
    timescales longer than a day, which are removed before searching
    for transits, are shown in blue, offset by 0.005 for clarity. See
    text for further details. \label{exlc:1}}
\end{figure*}

This letter presents a global assessment of the noise properties of
the CoRoT planet-finding channel data to date, from a
transit detection perspective, based on data collected over the first
14 months of the satellite's operations.

A preliminary discussion of CoRoT's photometric performance from an
instrumental point of view was given by \citet{abb+09}. In this work,
we start from the science grade data as released to the CoRoT
community and, one year later, to the public. We focus specifically on
the `white' ($\sim 300$--$1000$\,nm) light curves collected in the
planet-finding channel of CoRoT.

We detail the pre-processing steps taken to minimise instrumental and
astrophysical signals which can impede transit detection, and discuss
the noise on transit timescales after pre-processing. The motivation
for doing this is two-fold. First, it enables us to check whether the
CoRoT planet-finding data meets its pre-launch specifications, and
provides the first estimate of correlated noise levels in space-based
time series based on a large dataset. Second, it provides us with a
means of estimating the detection threshold for planetary transits as
a function of stellar magnitude.

This is the first in a series of three papers: based on the
photometric performance detailed here and on the results of the
transit detection and ground-based follow-up program to date:
\citet{paf09} presents an assessment of the overall CoRoT detection
threshold, which in turn serves as the input for \citet{fpa09}'s
attempt at a preliminary statistical interpretation of the CoRoT
planet catch to date.

\section{Light curve pre-processing}

The CoRoT N2 (science-grade) data contain a number of artifacts and
signals of astrophysical origin other than transits, which must be
removed or reduced before the transit search can proceed. In this
section, we detail the steps taken to do this, since it is the
properties of the pre-processed light curves which determine the
detectability of transits. In practice, of course, the CoRoT data are
and will be analysed by different teams employing a variety of
methods, which will no doubt be improved upon in the future. However,
the output of the method presented here is, to the best of our
knowledge, representative of the quality of the light curves used to
detect transits in CoRoT data to date.

To illustrate the effects to be removed, three typical CoRoT N2 light
curves for bright $(V<12.5)$ stars are shown in Fig.~\ref{exlc:1}
(in grey). All were taken from the second CoRoT long run
(150\,d), LRa01, during which the satellite was pointed in the
direction of the Monoceros constellation.

Most pro-eminent are upward outliers, due to the increased background
during the satellite's quasi-periodic crossings of the South Atlantic
Anomaly (SAA), and a long term decay assumed to be of instrumental
origin. Although data points collected during the SAA crossings are
flagged by the N2 pipeline, not all of these data points are
outliers. To maximise the duty cycle of the final light curves, we
apply a short baseline iterative non-linear filter with $3$-$\sigma$
outlier rejection to identify and reject outliers. The non-linear
filter consists of a 5-point boxcar filter followed by a median filter
whose baseline was chosen to correspond to one hour (see
\citealt{ai04} for further details of this type of filter).

There are also, though at a much lower level, downward outliers
associated with the satellite's entry into and exit from the Earth's
shadow, which introduces a temporary loss of pointing accuracy (these
are most clearly visible in the third example on Fig.~\ref{exlc:1}
around date 2\,885). These losses of pointing accuracy generally
affect a single data point, which is flagged by the N2 pipeline. We
use these flags to clip out the affected data. The example light
curves after clipping are shown in black on Fig.~\ref{exlc:1}. 

For light curves which are binned on board to 512\,s
sampling\footnote{On-board exposure times are 32\,s, but most data are
  rebinned to 512\,s by averaging sets of 16 exposures on board. Where
  this is not the case, we perform the rebinning ourselves so as to
  work with a data set with approximately uniform sampling, though we
  use medians rather than averages to minimise sensitivity to
  outliers.}, some points which appear affected by eclipse entry and
exit events are not flagged. For short (21d) runs, usually taken
around equinoxes when theses events are most pro-eminent, we fold the
light curves at the satellite orbital period and automatically
identify the remaining data points affected. It is not easy to do the
same thing for the longer runs, as the magnitude of these events and
their timing within the satellite's orbit evolve over timescales of
months, so some minor effects of the Earth eclipses remain in those
light curves.

All light curves share a long term downward trend, presumably due to
some kind of instrumental decay or gradual pointing drift. In
addition, most light curves show variations on timescales of days to
weeks, which are generally interpreted as resulting from rotational
modulation and evolution of active regions on the surface of the
star. We estimate the long-term component of the light curve
variations using a 1-day iterative non-linear filter (where the median
filter has a 1-day baseline and the other parameters are as
before). This estimate, shown in blue in Fig.~\ref{exlc:1}, is
subtracted from the clipped light curve, and the resulting detrended
light curve is used to perform the transit search.

Most light curves also contain one or more sudden jumps, which are
often followed by an exponential decay, though in some cases the decay
is delayed and sudden rather than exponential. These so called 'hot
pixel events' are caused by high-energy particles impacts on the
detector, which temporarily alter the sensitivity of one or more
pixels (see \citealt{prl+08}). Hot pixels leave local high-frequency
residuals in the detrended light curves which can be a source of false
alarms in the transit search, though they do not obviously affect the
global detrended light curve scatter.

In some cases, shorter timescale variations are also visible; these
can be periodic or quasi-periodic, as expected for pulsations or residual
satellite pointing jitter, or stochastic, as expected for e.g.\ surface
granulation. This type of variability, occuring on typical transit
timescales, cannot be filtered out without affecting a potential
transit signal.

\section{Noise estimates}

\begin{figure*}
  \centering
  \includegraphics[width=\linewidth]{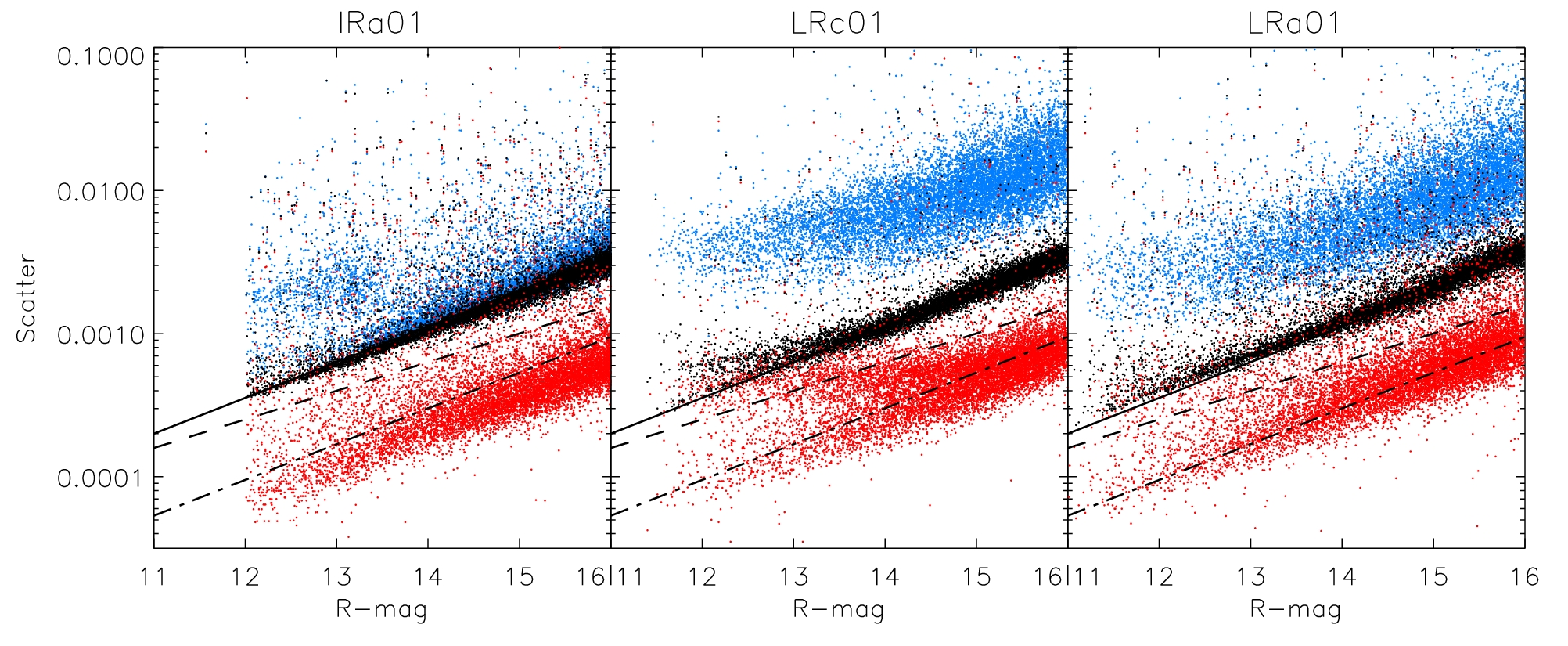}
  \caption{Noise level as a function of $R$-magnitude for IRa01, LRc01
    and LRa01. The black points represent the point-to-point scatter
    for each light curve. The solid line marks the lower enveloppe of
    these points, empirically approximated as a powerlaw with slope
    0.25 and value 0.2\,mmag at $R=11$. This can be directly compared
    with the theoretical (source) photon noise level, shown as the
    dashed line. The red points show the correlated noise level on
    2\,h timescales. This can be directly compared to the uncorrelated
    (white) noise on the same time-scales, shown as the dash-dotted
    line (which is obtained by scaling the solid line down by a factor
    $\sqrt{7200/512} \approx \sqrt{14}$). Also shown in blue is the
    light curve scatter on timescales of a day and longer, including
    the effect of stellar variability, instrumental decay / drift and
    hot pixels. \label{fig:rms}}
\end{figure*}

As shown by \citet{pzq06}, what determines the detectability of
transits is the noise level over a typical transit duration, which is
generally not equal to what one might infer from the point-to-point
scatter of the light curve, due to the presence of correlated noise on
hours timescales. We therefore evaluate the scatter of each light
curve after detrending over both 512\,s (point-to-point) and 2\,h
(CoRoT is sensitive to transits with periods up to a few weeks, which
last approximately 2\,h). The latter estimate is obtained by applying a
2\,h baseline iterative non-linear filter to the detrended light
curve. In both cases, we evaluate the scatter as $\sigma=1.48
\times{\rm MAD}$, where MAD is using the median of absolute deviations
from the median, an estimate of the standard deviation that is robust
to outliers \citep{hmt83}.

If the noise was entirely white, the ratio between the two estimates
(point-to-point and over 2\,h) should correspond to the square root of
the number of data points in a 2\,h interval (in this case 14). The
quadrature difference between the measured noise level on 2\,h
timescales and the value expected if the noise were white provides an
estimate of the component of the noise which is correlated on 2\,h
timescales. For comparison, we also evaluate the scatter on long
timescales from the light curve component which is subtracted in the
detrending process.

\begin{figure*}
  \centering
  \includegraphics[width=\linewidth]{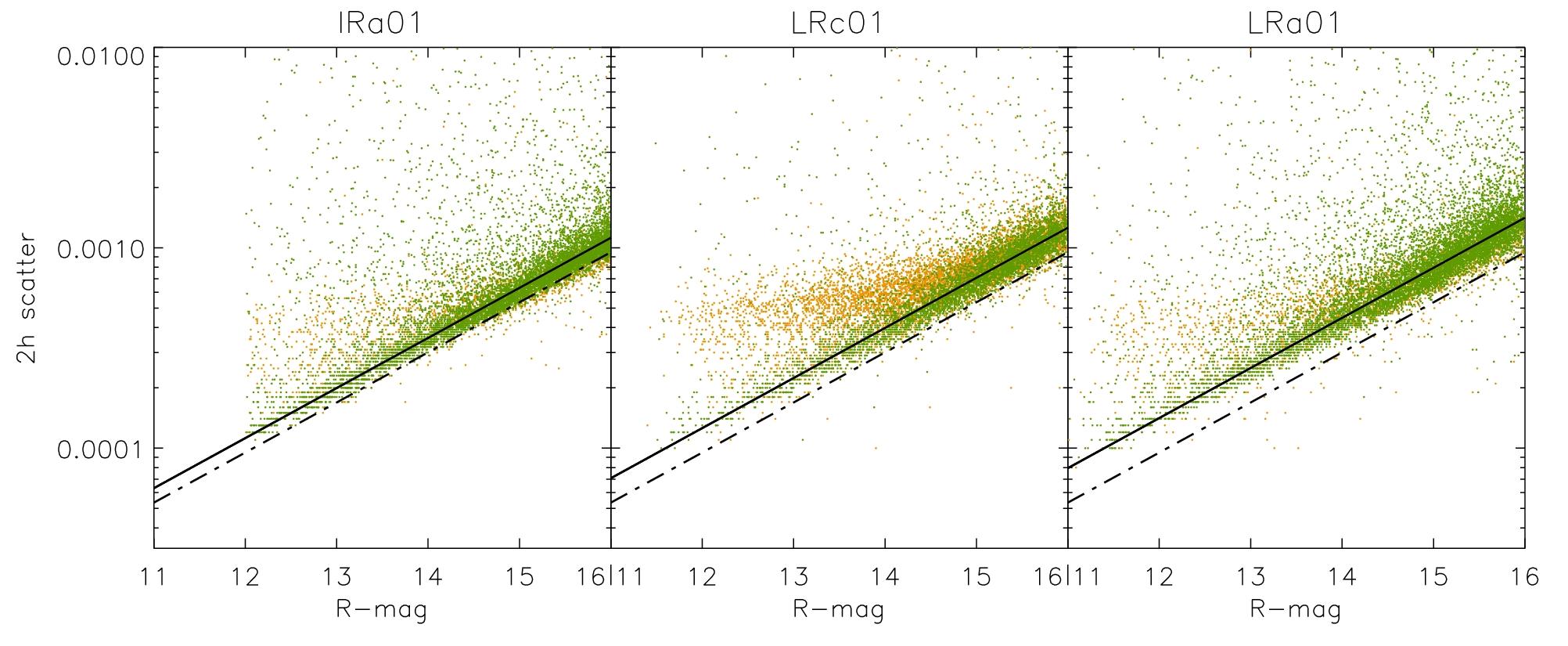}
  \caption{Median noise over 2\,h (including both correlated and
    uncorrelated components) as a function of $R$ magnitude for IRa01,
    LRc01 and LRa01. Green dots correspond to stars identified as
    likely dwarfs from the $R$, $B-V$ colour-magnitude diagram (see
    Fig.~\protect\ref{fig:cmds}), orange dots to stars identified as
    likely giants. The solid line represents the empirical formula of
    Eq.(\ref{eq1}), and the uncorrelated (white) component of the
    noise on 2\,h timescales is shown for comparison as the
    dash-dotted line. \label{fig:rms2}}
\end{figure*}

\begin{figure*}
  \centering
  \includegraphics[width=\linewidth]{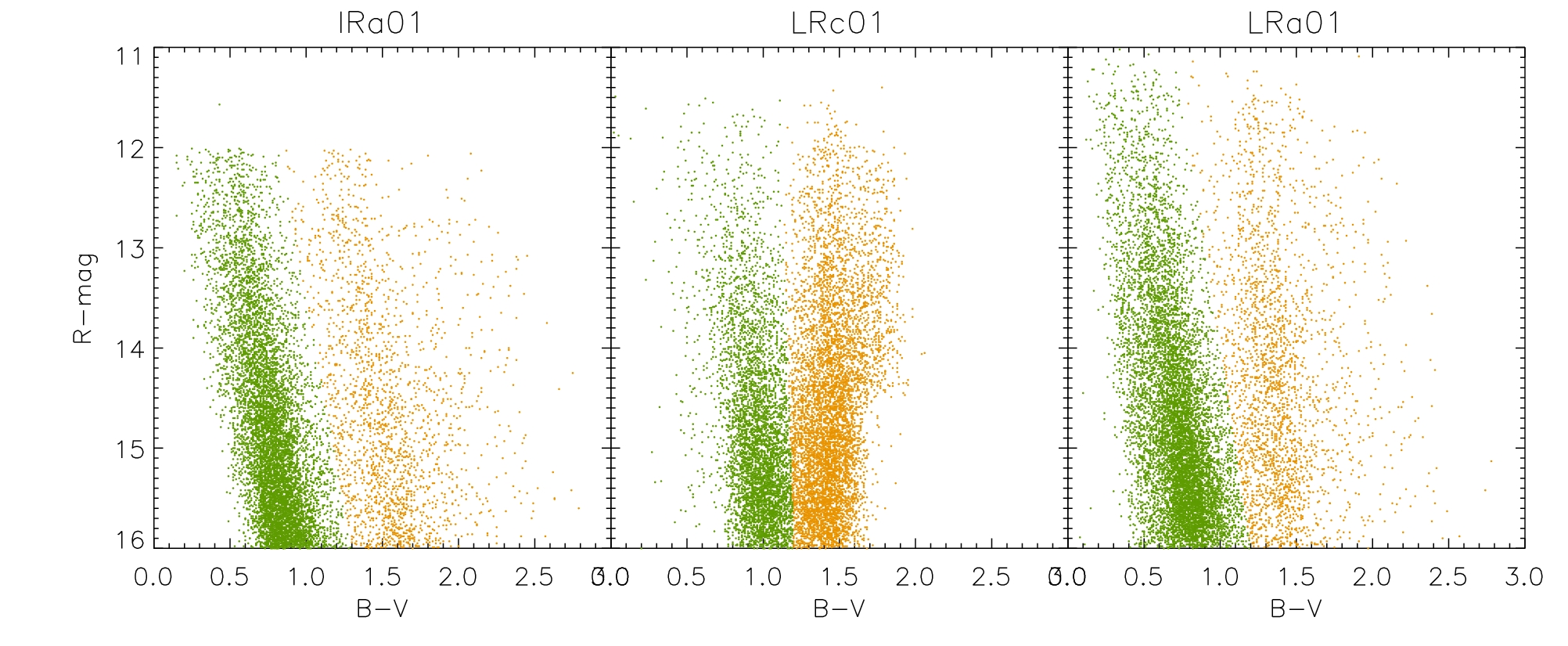}
  \caption{$R$-magnitude versus $B-V$ colour of all CoRoT targets in
    each field (data from Exodat, Deleuil et al.\ submitted). Green
    dots correspond to stars identified as likely dwarfs, orange dots
    to likely giants. The cut between dwarfs and giants was
    empirically set as a straight line running from $B-V = 0.7$, 1.1
    and 0.7 at $R=11$ to $B-V=1.3$, 1.2 and 1.2 at $R=16$ for IRa01,
    LRc01 and LRa01 respectively. The larger fraction of giants in
    LRc01 is clearly visible. \label{fig:cmds}}
\end{figure*}

This was done for the CoRoT initial run IRa01 (60d in the Galactic
Monoceros direction) and the first two long runs (LRc01 in the Aquila
direction, and LRa01). The results are shown in Figs.~\ref{fig:rms}
and \ref{fig:rms2}. Each data point in both figures is a MAD-estimated
scatter measurements for one star over the entire
run. Fig.~\ref{fig:rms} shows the noise per 512\,s exposure (black),
compared to the noise on daily and longer timescales (blue), and the
correlated noise over 2\,h (red). Fig.~\ref{fig:rms2} shows the total
noise over 2\,h, for stars identified as likely dwarfs (green) and
giants (orange) from $R$ versus $B-V$ colour magnitude diagrams, which
are shown in Fig.~\ref{fig:cmds}. We note that the colour-magnitude
cuts we adopted are an over-simplification, and in particular also
select some K-dwarfs as giants, but that the global inferred fractions
of giants in each run (20\% in IRa01, 58\% in LRc01 and 22\% in LRa01)
are in agreement with more sophisticated estimates based on
multi-colour photometry and, where available, spectroscopy (see
Barbieri et al.\, Gandolfi et al.\ in prep.).

\section{Discussion}

The lower envelope of the point-to-point scatter (solid line in
Fig.~\ref{fig:rms}) is a factor 2 above the photon noise (dashed line)
at the bright end, and rises somewhat more steeply (slope 0.25 in log,
versus 0.2 for photon noise). However, the correlated noise on transit
timescales (red points) is comparable to or slightly smaller than the
white noise on 2\,h timescales (dot-dashed line). The empirical
relations between total 2\,h noise and $R-$magnitude for dwarfs, shown
as solid lines on Fig.~\ref{fig:rms2}, are given by:
\begin{equation}
  \label{eq1}
  \log \sigma_{\rm 2h} = 0.25 \, R + z
\end{equation}
where $z=-6.95$ for IRa01, $-7.0$ for LRc01 and $-7.05$ for IRa01,
indicating a gradual degradation of the photometric performance over
time, which may be associated with the increase in incidence of hot
pixel events. Again, the slope of this relation is slightly above that
expected for source photon noise.

The pre-launch specification for CoRoT's photometric performance is
$7.7\cdot 10^{-4}$ in 1\,h down to $V=15.5$ \citep{abb+03}. This
specification was designed to ensure the detectability of transits of
planets down to approximately $1.5\,R_{\rm Earth}$ in short-period
($<1$ week) orbits, assuming that noise averaged out in a white manner
over timescales longer than 1\,h. Thus it corresponds to $5.5\cdot
10^{-4}$ in 2\,h, while $V=15.5$ corresponds approximately to $R=15$
for a typical CoRoT dwarf target. Eq.~(\ref{eq1}) gives values of 6.3,
7.1 and $7.9 \cdot 10^{-4}$ at $R=15$, close to, but slowly receding
from, the pre-launch specification.

Although this level of photometric performance is already a vast
improvement on any previously available long-baseline photometry of
large samples of stars, there clearly remains scope for further
improvement. The behaviour of the correlated noise with magnitude
suggests it may dominated by effects such as satellite pointing jitter
(as the aperture masks are smaller for fainter stars), and imperfect
background correction. Improvements to the light curve generation,
including e.g.\ a new jitter correction based on improved PSF
modeling, and a posteriori ensemble light curve analysis, have
recently been tested on CoRoT data (see Fiahlo et al., Mazeh et al.\
in prep.), and these will hopefully lead to improved performance in
the short to medium term.  The importance of developing an effective
method for the treatment of light curve discontinuities, more
sophisticated than the detrending applied here, is rapidly increasing,
as visual examination of light curves from LRa01 indicates that almost
every light curve is affected by at least one discernible
discontinuity, compared to approximately one in four in IRa01. Such
techniques are under development, though it is challenging to apply
them in an automated fashion without unwittingly affecting other
signals in the light curves.

\begin{table}
  \caption{Percentage of light curves more variable over 2\,h 
    timescales than 1.5 times the relation of 
    Eq.~(\protect\ref{eq1}), for each run, as a function of magnitude
    (see text for details). Numbers in brackets show the fraction 
    of variables when considering only stars identified as likely 
    dwarfs from the $R$ versus $B-V$ colour-magnitude diagram.
    No data is shown for magnitude bins containing fewer 
    than 20 objects.}
  \label{tab:rms}
  $$ 
  \begin{tabular}{lrlrlrl}
    \hline\hline
    \noalign{\smallskip}
    R{\rm -mag} & \multicolumn{6}{c}{\rm \% variable stars} \\
    & \multicolumn{2}{c}{\rm IRa01} & \multicolumn{2}{c}{\rm LRc01} & 
    \multicolumn{2}{c}{\rm LRa01} \\
    \noalign{\smallskip}
    \hline
    \noalign{\smallskip}
    11.0 & \multicolumn{2}{c}{-} & \multicolumn{2}{c}{-} &   60 & (63) \\
    11.5 & \multicolumn{2}{c}{-} &   67 & (35) &   62 & (47) \\
    12.0 &   56 & (38) &   68 & (36) &   46 & (42) \\
    12.5 &   42 & (33) &   58 & (27) &   39 & (31) \\
    13.0 &   35 & (30) &   54 & (26) &   31 & (33) \\
    13.5 &   28 & (24) &   37 & (17) &   22 & (25) \\
    14.0 &   20 & (21) &   20 & (14) &   17 & (22) \\
    14.5 &   17 & (19) &    9 &  (9) &   14 & (19) \\
    15.0 &   11 & (13) &    5 &  (8) &   11 & (15) \\
    15.5 &    9 & (11) &    5 &  (6) &   11 & (13) \\
    16.0 &    8 &  (9) &    5 &  (8) & \multicolumn{2}{c}{-} \\
    \noalign{\smallskip}
    \hline\hline
  \end{tabular}
  $$ 
\end{table}

Although we now have a simple relationship between noise on
transit timescales and magnitude for dwarf stars, which can be used to
estimate transit detection thresholds \citep{paf09}, for a statistical
interpretation of the CoRoT planet catch \citep{fpa09}, it is also
important to know what fraction of the stars deviate significantly
from this relation. Table~\ref{tab:rms} lists the percentage of light
curves, in each run, whose scatter over 2\,h timescales is more than
1.5 times above the relation given by Eq.~(\ref{eq1}). At the bright
end, 50\% or more of the light curves show significant variability on
transit timescales, rising to 70\% in LRc01, while the fraction of
drops to a few percent at the faint end.

Interestingly, because it is closer to the direction of the Galactic
centre, LRc01 has a significantly larger fraction of giants (see
Fig.~\ref{fig:cmds}). Fig.~\ref{fig:rms2} clearly shows that a large
fraction of stars identified as likely K-dwarfs or giants from the $R$
versus $B-V$ colour-magnitude diagram vary at the 0.5\,mmag level on
2\,h timescales, whereas no such trend is seen in stars identified as
likely F or G dwarfs. In a detailed Fourier domain study of the light
curves of a sample of red giants observed by CoRoT, \citet{kwb+09}
found significantly enhanced variability on timescales of a few hours
compared to the Sun, but we confirm here that this is a systematic
effect, clearly visible over thousands of stars spanning a range of
colours. We note this enhanced variability was expected from 3--D
simulations of granulation \citep{fre01,sl05}, which indicate that
granulation cell size -- and hence the strength and timescale of the
photometric signature of granulation -- increase significantly as
surface gravity decreases.

\section{Conclusions}

We have evaluated the noise per exposure and on 2\,h timescales for
the first three CoRoT observing runs lasting more than a month. We
find that the photometric performance on transit timescales is close
to the pre-launch specification, with a level of correlated noise at
least an order of magnitude below that obtained from the ground. The
observed performance is sufficient to detect transits of planets of a
few Earth radii in short period ($<3$\,d) orbits around bright stars.

However, there is scope for further improvement, since the noise level
per exposure exceeds the photon noise by a factor 2 to 3 and the level
of correlated noise, although low, is non-zero. The slight degradation
of the performance from IRa01 to LRa01 indicates that the impact of
hot pixels is gradually increasing, so that improving the treatment of
these artifacts is becoming increasingly important. 

We also show that giants -- tentatively identified from the $R$ versus
$B-V$ colour-magnitude diagram -- constitute a significant fraction of
the CoRoT targets (from 20\% for Monoceros runs to 50\% for Aquila
runs). Giants are in any case less than ideal targets for transit
surveys due to their large radii, but we also find that they tend to
be variable at the 0.5\,mmag level on transit timescales, which we
interpret as a granulation effect.

Finally, one aspect which has not been touched upon here so far, but
will be explored in more detail in a forthcoming paper, is the fact
that the global variability levels on timescales of a day and longer
(blue points in Fig.~\ref{fig:rms}) are relatively high: very
few CoRoT light curves have dispersions below 1\,mmag on those
timescales. For comparison, we measured in the same way the scatter of
the total irradiance variations of the Sun from SoHO/VIRGO/PMO6 at
times of minimum and maximum activity (see \citealt{afg04} and
references therein for details of this dataset), finding 0.07 and
0.3\,mmag respectively. However, to quantify this tentative but
interesting result further requires discrimination between the effects
of instrumental decay, hot pixels, satellite pointing jitter and true
stellar variability, which is beyond the scope of this paper.

\begin{acknowledgements}
  HD acknowledges support from grant ESP2007-65480-C02-02 of the
  Spanish Science and Innovation ministry, the German CoRoT team (TLS
  and Univ.\ Cologne) from DLR grants 50OW0204, 50OW0603, and
  50QP0701, and SZ from the Israel Science Foundation -- Adler
  Foundation for Space Research (grant No. 119/07).
\end{acknowledgements}

\bibliographystyle{aa}
\bibliography{CoRoTNoise}

\end{document}